# Unveiling the Glass Veil: Elucidating the Optical Properties in Glasses with Interpretable Machine Learning

Mohd Zaki[1], Vineeth Venugopal[1], R. Ravinder[1], Suresh Bishnoi[1], Sourabh Kumar Singh[1], Amarnath R. Allu[2], Jayadeva[3,*], N. M. Anoop Krishnan[1,4,*]

[1]Department of Civil Engineering, Indian Institute of Technology Delhi, Hauz Khas, New Delhi 110016, India
[2] CSIR-Central Glass and Ceramic Research Institute, Kolkata 700032, India
[3]Department of Electrical Engineering, Indian Institute of Technology Delhi, Hauz Khas, New Delhi 110016, India
[4]Department of Materials Science and Engineering, Indian Institute of Technology Delhi, Hauz Khas, New Delhi 110016, India

[*]Corresponding authors: Jayadeva (jayadeva@iitd.ac.in), N. M. A. Krishnan (krishnan@iitd.ac.in)

**Abstract**
Due to their excellent optical properties, glasses are used for various applications ranging from smartphone screens to telescopes. Developing compositions with tailored Abbe number ($V_d$) and refractive index ($n_d$), two crucial optical properties, is a major challenge. To this extent, machine learning (ML) approaches have been successfully used to develop composition–property models. However, these models are essentially black-box in nature and suffer from the lack of interpretability. In this paper, we demonstrate the use of ML models to predict the composition-dependent variations of $V_d$ and n at 587.6 nm ($n_d$). Further, using Shapely Additive exPlanations (SHAP), we interpret the ML models to identify the contribution of each of the input components toward a target prediction. We observe that the glass formers such as $SiO_2$, $B_2O_3$, and $P_2O_5$, and intermediates like $TiO_2$, PbO, and $Bi_2O_3$ play a significant role in controlling the optical properties. Interestingly, components that contribute toward increasing the $n_d$ are found to decrease the $V_d$ and vice-versa. Finally, we develop the Abbe diagram, also known as the "glass veil", using the ML models, allowing accelerated discovery of new glasses for optical properties beyond the experimental pareto front. Overall, employing explainable ML, we discover the hidden compositional control on the optical properties of oxide glasses.

**Introduction**
Since the discovery of glass manufacturing in 3000 B.C. and the use of clear glass by the Romans for architectural purposes in 100 A.D., glasses have emerged as one of the important materials in human history [1,2]. With the emergence of eye-glasses in the 13[th] century, the discovery of microscope in the 16[th] century, followed by the advancements in the fields of astronomy due to telescope in the 17[th] century [3], and camera lens in 19[th] century [4], the need for glasses with tailored optical properties has been ever-increasing [5]. Due to the ability of nearly every element in the periodic table to form glass, when mixed with glass formers such as $SiO_2$, $P_2O_5$, $B_2O_3$, there is a possibility of making an astronomical number of glass compositions[6]. However, experimentally making and testing the properties of such a large number of compositions will be extremely prohibitive in terms of time and cost. Further, not all of these glasses may be of interest for optical applications. As such, designing novel optical glasses require the development of reliable composition–property models for predicting the optical properties of glasses [7].

Recently, researchers have started employing data-driven approaches, such as machine learning (ML) models, to predict the properties of oxide glasses [8–20]. Recent studies have developed largescale composition–property models, accommodating ~9 properties and ~40 compounds, for predicting the properties of oxide glasses [13,21]. Ravinder et al.(2020) presented the concept of glass selection charts (GSCs) to develop novel glasses, which allows the selection of components for designing glasses with tailored properties [21]. Note that these GSCs are very similar to the Abbe diagram, also known as the "glass veil", which shows the variation of Abbe number ($V_d$) and refractive index at 587.6 nm ($n_d$) for a variety of optical glasses. Abbe diagram is an extremely useful tool to obtain glasses with targeted $V_d$ and $n_d$, which can then be used for specific applications. For example, in optical applications where multiple lenses are needed for obtaining achromatic pairs, the question of which oxides to make different glasses becomes tougher because researchers need to obtain optimized glass compositions for multiple glasses in that case.

Tokuda et al. (2020) used statistical methods to predict optical properties of just 879 glass compositions, which is very less given the glass-forming ability large number of oxides [22]. Cassar et al. (2020) presented an ML-aided optimization paradigm for optical glasses design [23]. However, despite giving good results, these ML models failed to explain the model output, thus does not give any insight to researchers why a prediction has been made. In the ML domain, predictions of the models are inexplainable because of their highly complex nature. Recent studies on explainable ML have placed emphasis on interpreting the models using additional ML tools [24,25]. Some attempts toward this direction have been attempted in glass science as well by applying explainable ML algorithms for predicting glass transition temperatures [26]. However, none of these attempts could deconvolute the contribution of a given component toward a target prediction, thereby allowing direct interpretation of the model. It is possible to address this challenge using Shapely Additive exPlanations (SHAP), a concept developed based on the Shapely values from game theory. Recently, SHAP has been used in a wide range of fields including biomedical engineering [27], civil engineering [28], and materials science to interpret the ML "black-box" models [29].

In this paper, we develop ML models for the two most important optical properties of glasses: $V_d$ and $n_d$. The ML models are trained on more than 10,000 unique compositions for both properties, with 47 and 49 oxide components available for $V_d$ and $n_d$, respectively, details of which are provided as supplementary material. The trained models are used to populate the ternary diagrams for two distinct glass families of crown and flint to demonstrate the models' ability to learn the physics governing these two optical properties. Further, using SHAP [25,30], we identify the oxides which have a positive/negative impact on the optical properties. Interestingly, we observe that the oxides which contribute to an increase in the $n_d$ value affect $V_d$ negatively, that is, reduce the property value and vice-versa. Finally, using the ML models, we develop the Abbe diagram or the glass veil, which will allow the researchers to develop novel glass compositions for optical applications.

**Methodology**
**Data preparation**
The dataset of $n_d$ and $V_d$ of oxide glasses are obtained from curated databases such as the International Glass Database (INTERGLAD V7.0) and SciGlass. These are comprehensive sources of glass composition and property data along with the relevant references from which the information has been obtained [31]. Since these datasets have been manually compiled, there are some inconsistent entries, for example, compositions not summing up to 100 mol%, duplicate or missing entries. To address these issues, only those compositions which summed up to 100 mol% were selected. Further, duplicate entries were deleted. Occasionally, entries were having multiple property values corresponding to the same composition. In this case, the values lying beyond ±2.5% of the mean property values for the given composition were discarded as outliers. Further, the mean of remaining values was used as the representative property value for the given composition. Finally, to ensure representative data in both training and test set, only those oxide components present in at least 30 or more glass compositions were selected. This protocol resulted in ~14,000 unique compositions and 49 oxide components for $n_d$ and ~11,000 unique compositions and 47 oxide components for $V_d$.

**Model training**
The cleaned data was split as 80:20 to obtain training and test sets. For model training, the oxide components were taken as input features and optical properties ($n_d$ and $V_d$) as output for training ML models. Each row in the input dataset has molar percentage corresponding to each oxide component present in the glass. Absent oxide components are assigned a value of zero. Note that the test set was kept as a holdout set and was never used in the training process to ensure reliable model testing. The training set was further subjected to four-fold cross-validation, thus providing 60:20:20, training, validation, and test sets. The mean and standard deviation of the training set is used to scale the data before starting model training. PyTorch, an open-source ML library, was used for developing a feedforward neural network (NN). Inbuilt functions were used to implement batch normalization and drop out during hyperparametric optimization, ensuring that the models do not overfit during training.

Two separate models were trained for predicting the two properties. A two hidden layer neural architecture with 19 hidden layer units in each layer, and a ReLU activation function, was used. This architecture has been obtained after a grid search, where the number of hidden layers was varied from one to three, and the number of neurons in each layer was varied from 1 to 30. Note that dropout was not used while arriving at the optimal hidden layer architecture. For models with more than one hidden layer, the number of neurons was kept the same in all the layers. To train the final NN, the mean squared error (MSE) was minimized using the Adam optimizer [32]. The dropout probabilities of 0.1 and 0.2 were used while training the models for $n_d$ and $V_d$, respectively, to prevent overfitting. In addition to dropout, L2 regularization was also enforced for optimizing the NN weights to ensure the fitness of the models. The final models obtained were tested using the test set and used for further analysis to make the glass selection chart (GSC). The python packages NumPy [33], pandas [34] and matplotlib [35] have been used for preprocessing and visualization of the dataset and results. The code for implementing SHAP can be found at the following open-source GitHub repository: https://github.com/slundberg/shap

**Shapely Additive exPlanations (SHAP)**
Being a game theory concept, Shapely values depend upon two attributes, namely, game and players. Here, the game is analogous to predicting the $V_d$ and $n_d$ of a glass composition while the input oxide components are equivalent to the players. Shapely Additive exPlanations (SHAP) [25] is a recent technique developed, based on Shapely values, to explain the contribution of the individual features towards a prediction. SHAP value corresponding to each oxide in the model predictions can be defined as

$$SHAP_{oxide|c} = \sum_{set:oxide \in set}[|set| \times \binom{F}{|set|}]^{-1} \times [predict_{set}(x) - predict_{set \setminus oxide}(x)]$$
Eq (1)

where, c is the glass composition, set is the power set of all features in which oxide is present, |set| is the number of oxides in the subset of the set under consideration, F is the total number of oxides present in glass composition c, $predict_{set}$ is the model predictions using oxides present in the set as input, $predict_{set\backslash oxide}$ is the model predictions using oxides except the oxide for which SHAP value is being calculated, x is the subset of c corresponding to the features in the set having the oxide component of interest, prediction for the null set is the mean of the model output for the training set.

To demonstrate how SHAP can be applied for interpreting the NN model prediction, consider a ternary glass composition C1 having oxide features as f1, f2 and f3, for example, CaO, $Na_2O$, and $SiO_2$. The power set of oxides for C1, P(C1), will be { { }, {f1}, {f2}, {f3}, {f1,f2}, {f2,f3}, {f1,f3}, {f1,f2,f3} } and the total number of oxides, F will be 3. Hence, the SHAP value of feature f1 in C1 can be calculated using the above equation, where set will be { {f1}, {f1,f2}, {f1,f3}, {f1,f2,f3} }, that is, only those sets where f1 is present. Note that the underlying assumption in SHAP is features contribute independently to the final prediction.

**Results**

The cleaned dataset for $n_d$ consists of 14269 unique glass compositions and a total of 49 unique oxide components, including 532 are binaries, 2279 are ternaries, 684 are quaternaries, and 845 are five component glasses (see Supplementary Materials for details). Similarly, the cleaned dataset for $V_d$ consists of 11335 unique glass compositions and 47 unique glass components, including 287 are binaries, 1394 are ternaries, 409 are quaternaries, and 684 are five component glasses. Figure 1 shows the experimental and predicted values of $n_d$ and $V_d$ as a heat map for the test dataset. The $R^2$ scores for training, validation, and test for $V_d$ and $n_d$ are (0.97, 0.97, and 0.95) and (0.98, 0.98, and 0.97), respectively, implying an optimal training procedure for the models. The clustering of data close to the 45° (y=x) line represents a good agreement between experimental and predicted values on the test set (see Figure 1). The heat map is used for representation owing to the large number of data points, where the color represents the number density of the points associated with $V_d$ in Figure 1(a) and $n_d$ in Figure 1(b). The error between the actual and predicted values are plotted as the probability density function in the inset. The narrow range of the shaded region, which represents the 95% confidence interval, confirms the performance of the models.

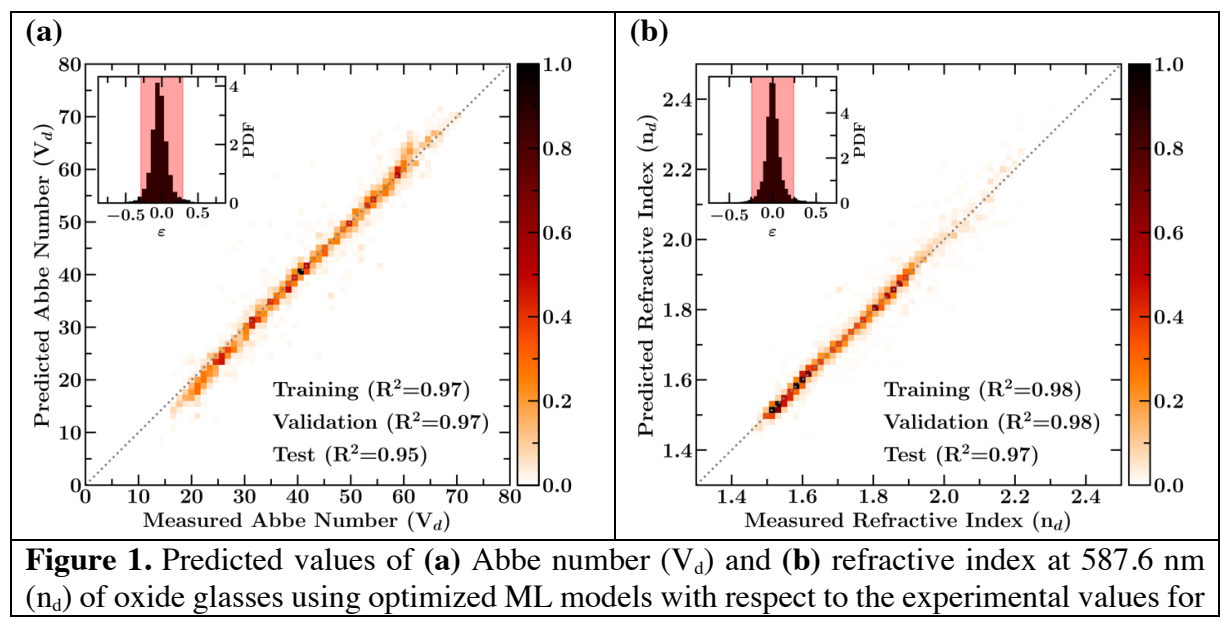

**Figure 1.** Predicted values of **(a)** Abbe number ($V_d$) and **(b)** refractive index at 587.6 nm ($n_d$) of oxide glasses using optimized ML models with respect to the experimental values for the test dataset. Note that the color represents the number of points per unit area associated with each property following the respective coloring scheme. The inset shows the error in the predicted values as a probability density function (PDF) with the shaded region representing the 90% confidence interval. Training, validation, and test $R^2$ values are also provided.

To further investigate the model performance, ternary plots for different crown and flint glass families have been shown in Figure 2. Note that crown glasses have high $V_d$ and low $n_d$, while flint glasses have low $V_d$ and high $n_d$ [36]. To demonstrate the ML model's capability in capturing the underlying physics governing both the optical properties, these models are used to predict the property values for all possible combinations of the ternary Barium borosilicate(BaBS) from the crown glass family (see Figs. 2(a) and (b)) and Potassium lead silicate(KPbS) from the flint glass family(see Figs. 2(c) and(d)). The experimental values plotted on top of the background (square markers) demonstrate that the predictions by the model are in good agreement with the experimental values. In addition, the model allows us to explore a larger domain, thereby suggesting new potential compositions that can serve as flint or crown glasses.

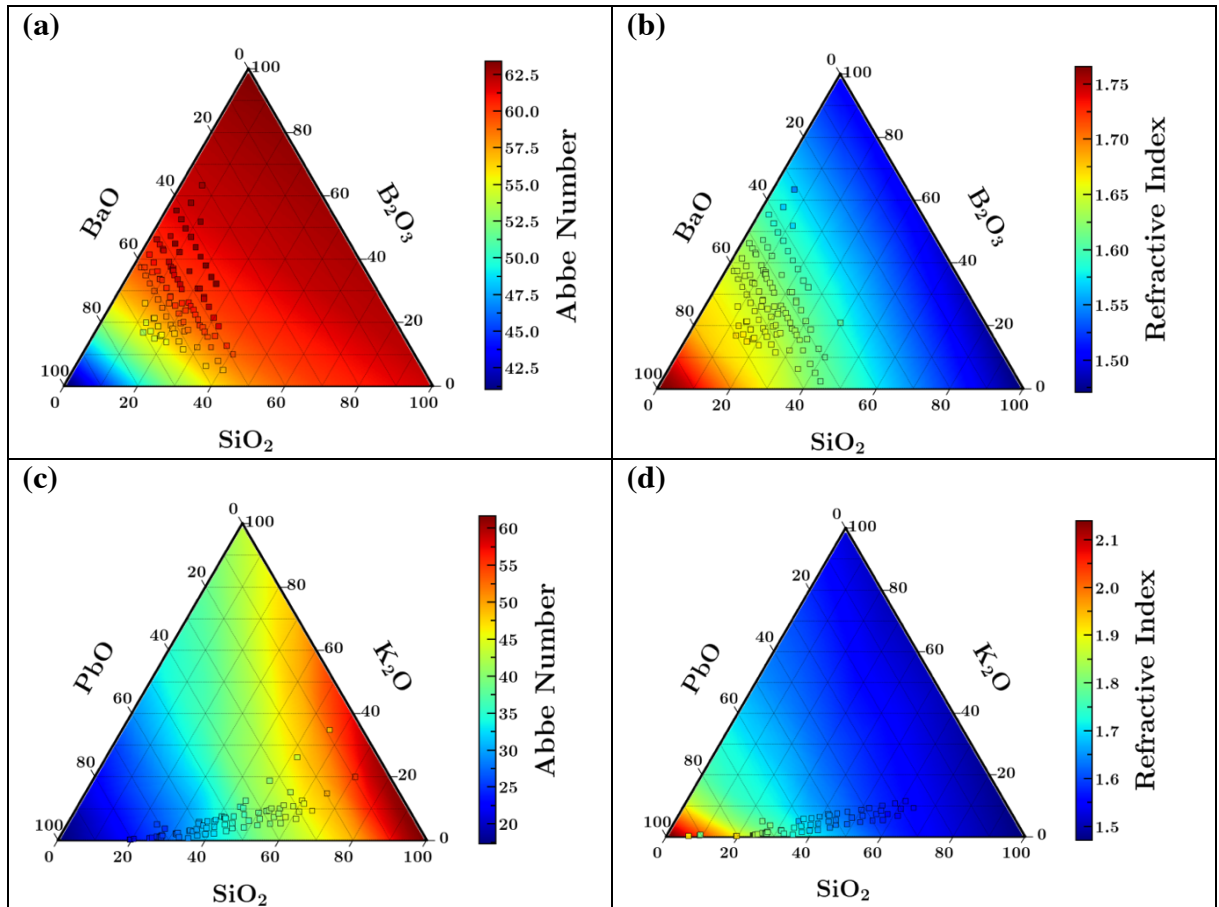

**Figure 2.** Ternary plot for **(a)** Abbe number ($V_d$) and **(b)** refractive index ($n_d$) of [BaO, $B_2O_3$, $SiO_2$] crown glass and **(a)** Abbe number ($V_d$) and **(b)** refractive index of ($n_d$)[PbO, $K_2O$, $SiO_2$]flint glass. Square markers represent experimental compositions with the color representing the property value from the database. The underlying heatmap represents the model predictions. The color bar corresponding to each subfigure shows the property range along with the coloring scheme.

Although the model is able to capture the details of the composition–property models, the model is too complex to be interpretable. Specifically, the NN models for $V_d$ and $n_d$ have highly complex structures with multiple hidden layers, and many fully connected hidden layer units. As such, these NNs are black-box models, making it unclear what the network has learned during the process. For problems in materials, in addition to predicting the output values, it is important to know how an oxide component affects the property values. This aspect is important to understand the underlying physics governing the composition–property relationship and correlate with known theories, thereby gaining increased confidence in the learning process and the ML model. To address this challenge, we use SHAP, which provides the contribution of each of the oxide components toward a given prediction (see Methodology).

As mentioned earlier, SHAP values can be computed for a given oxide component depending upon its composition corresponding to a prediction. The SHAP value suggests how much of a relative impact the given component has in the model prediction. If the SHAP value is high, the component has a high impact on the model prediction, and vice-versa. The mean absolute SHAP value is then obtained by computing the mean of the absolute SHAP values for a given oxide corresponding to all the glass compositions in which the oxide is present. Thus, if the mean absolute SHAP value of an oxide is high, the component has a higher impact on predicting the given property in an average sense. It should be noted that SHAP only deconvolutes the contribution of a component toward a prediction; it doesn't predict that an increase in a composition will lead to an increase in the output. Although the difference is subtle, it should be understood that SHAP is a post-mortem analysis of the contributions of components toward a given prediction.

Figure 3(a) shows the mean absolute SHAP value of the top 25 oxides governing $V_d$. We observe that the top two oxides which affect the $V_d$ significantly are $SiO_2$ and $TiO_2$. These are followed by $Nb_2O_5$, $P_2O_5$, $B_2O_3$, $Bi_2O_3$, and $PbO$ in the respective order. This list suggests that the above-mentioned components play a crucial role in controlling the $V_d$ of oxide glasses. Note that each of the components mentioned above can have a positive or negative effect on the $V_d$. Specifically, a given oxide component having a high SHAP value may increase or decrease the property. However, the mean absolute SHAP value does not convey this information.

To this extent, we focus on the SHAP values of the top 25 components corresponding to each of the predictions. Figure 3(b) shows the SHAP values of the top 25 oxides for the NN model corresponding to $V_d$. The color of the points represents the normalized molar concentration of an oxide component with the pink and blue colors indicating the higher presence and lower presence, respectively. We observe that the nature of the oxide's impact on the model output can be clearly understood from this plot. For example, although, $SiO_2$ and $TiO_2$ have high mean absolute SHAP values, the SHAP value of $SiO_2$ increases with an increase in its molar concentration while that of $TiO_2$ reduces with an increase in its molar concentration. This suggests that higher values of $SiO_2$ contribute toward increasing $V_d$, while that of $TiO_2$ contribute toward decreasing $V_d$, thereby providing a direct way to identify the components that contribute toward increasing or decreasing $V_d$, both qualitatively and quantitatively. The complete list of feature importance obtained using SHAP for both the properties is provided in the supplementary material.

From Figure 3(b), we classify the components into two categories, namely, the ones that contribute toward increasing and decreasing the values of $V_d$. Some of the oxides in the first category that is the ones contributing toward increasing $V_d$ are $SiO_2$, $P_2O_5$, $B_2O_3$, $BaO$, $La_2O_3$,

$Al_2O_3$, CaO, $K_2O$, $Gd_2O_3$, and SrO, in the respective order. Similarly, the ones contributing toward a decrease in $V_d$ are $TiO_2$, $Nb_2O_5$, $Bi_2O_3$, PbO, $TeO_2$, $Ta_2O_5$, $WO_3$, $ZrO_2$, and $Sb_2O_3$, in the respective order. These lists thus suggest the combination of oxides that should be chosen to arrive at a target $V_d$.

Similarly, Figure 4(a) represents the mean absolute SHAP values of the top 25 oxides for $n_d$. We observe that $SiO_2$ and $Bi_2O_3$ are the oxides significantly affecting the $n_d$ of a glass. The other components include $P_2O_5$, $Nb_2O_5$, $TiO_2$, $La_2O_3$, $B_2O_3$, and PbO, respectively. This list represents the components that contribute significantly toward the $n_d$. Figure 4(b) represents the SHAP value of the top 25 oxides for $n_d$. Interestingly, we observe that higher values of $SiO_2$ result in a reduction of the $n_d$ of glass, while the trend with $Bi_2O_3$ is the opposite. As in the case of $V_d$, we classify the components into the ones that contribute toward increasing or decreasing the $n_d$ values. The first class includes $Bi_2O_3$ $Nb_2O_5$, $TiO_2$, $La_2O_3$, PbO, and $TeO_2$, while the second class includes $SiO_2$, $P_2O_5$, $B_2O_3$, $Na_2O$, and $K_2O$, respectively, to name a few.

It is interesting to note that in the cases of both $V_d$ and $n_d$, network formers and intermediates (for example, $SiO_2$, $B_2O_3$, $P_2O_5$, $TiO_2$, and $Bi_2O_3$) contribute as major features governing the optical properties while network modifiers (for example, CaO, $Na_2O$, MgO, $Y_2O_3$, and ZnO) contribute relatively lower toward the property prediction. Interestingly, the oxides contributing significantly to increasing the $V_d$ value (for example, $SiO_2$, $B_2O_3$, $P_2O_5$) tend to decrease the $n_d$ of the glass and vice-versa (for example, $Bi_2O_3$, $TiO_2$, $Nb_2O_5$, PbO, and $TeO_2$). These observations are in agreement with the findings reported in the literature [36].

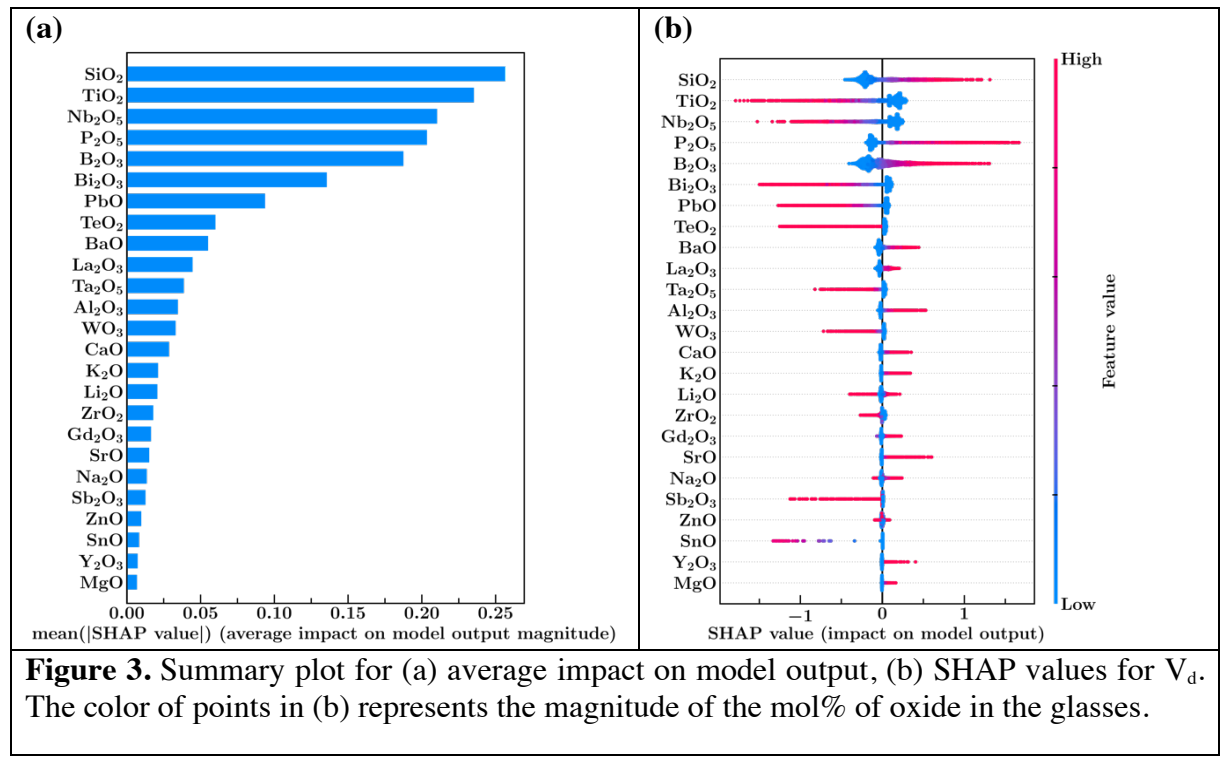

**Figure 3.** Summary plot for (a) average impact on model output, (b) SHAP values for $V_d$. The color of points in (b) represents the magnitude of the mol% of oxide in the glasses.

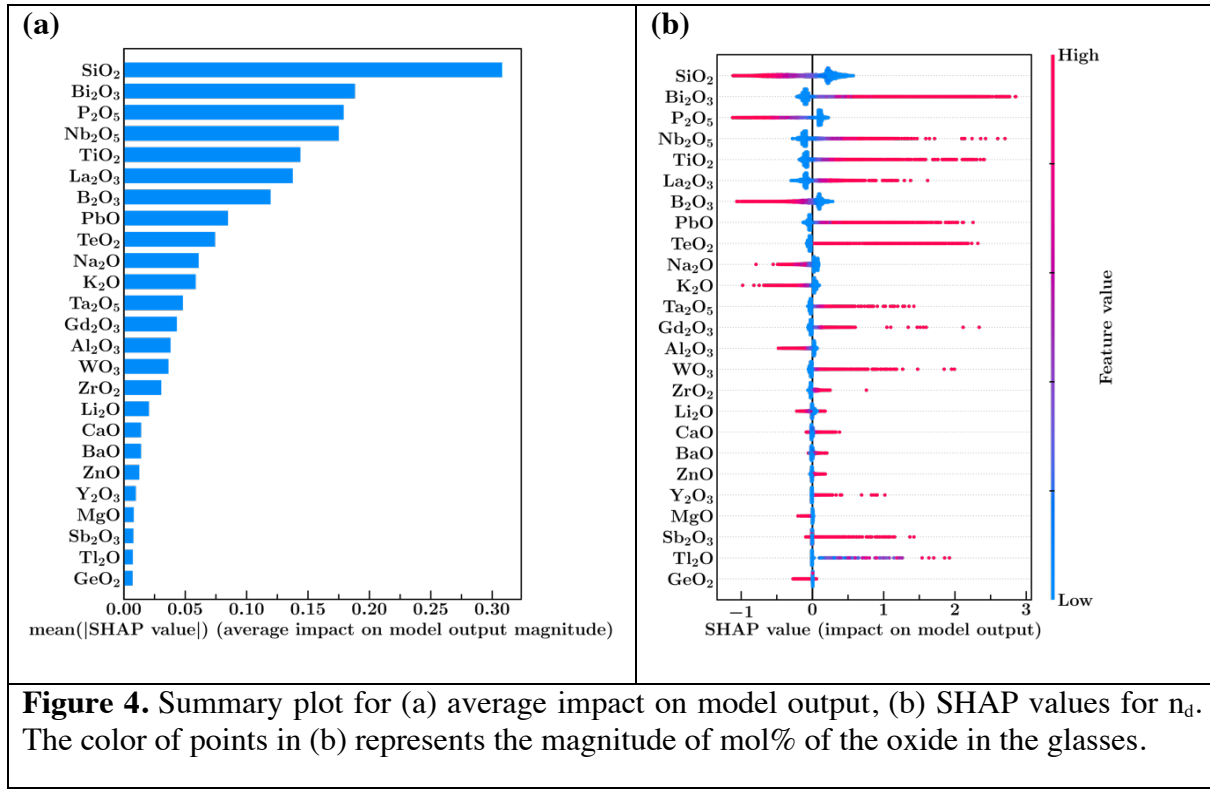

**Figure 4.** Summary plot for (a) average impact on model output, (b) SHAP values for $n_d$. The color of points in (b) represents the magnitude of mol% of the oxide in the glasses.

**Discussion**

Overall, we observe that SHAP allows a detailed interpretation of the ML models. Further, we find that the inferences from SHAP are in agreement with the empirical knowledge on the optical properties of glasses. For instance, flint glasses are characterized by high $n_d$ and low $V_d$. Figures 3 and 4 show that the components that contribute toward high $n_d$ and low $V_d$ are $Bi_2O_3$, $TiO_2$, $Nb_2O_3$, and PbO, to name a few. Interestingly, we find that most of the earlier flint glasses contained significant amounts of PbO, which were eventually replaced by $TiO_2$. Further, most of the flint glasses available are indeed comprising of these major components specified. Thus, SHAP values provide insights into the selection of components towards a targeted design.

To this extent, using the ML models developed, we present the Abbe diagram or glass veil in Figure5. In the Abbe diagram, the $V_d$ is taken on the x-axis in decreasing order, while $n_d$ is taken on the y-axis in increasing order. The color blobs in the graph correspond to various ternary, quaternary and quinary compositions for which both the optical properties are predicted using ML models. The points in the background of the colored blobs are taken from the experimental dataset of $n_d$ and $V_d$ databases on compositions for which both the optical properties are known. The colored region in the glass veil shows the following ternaries TiLaZr, BiTeNb, BaLaB, NaTiSi, PTiK, BaNbSi, NaPbSi, BaTaSi, BaSiB, BAlSi and KPbSi, quaternaries LaTiBSi and SiKZnZr, and quinaries BiGePbTeTl and MgPAlKB. These regions have industrial importance [36,37] and make it possible for ML models to explore the glass veil beyond the pareto front.

The glass veil in Figure 5 encompasses both crown and flint glasses. Traditionally, flint glasses contained lead, which contributed to pollution during the manufacturing and disposal phase. The overlapping of BiGePbTeTl blob, a flint glass, with the lead-free glass regions produced using ML models for TiLaZr and BiTeNb suggest the alternatives to mitigate such pollution. Note that titanium dioxide and zirconium dioxide are used to replace the lead oxide in the glass composition [38–41]. Crown glasses that find application in lenses have a low $n_d$ of around 1.5 and $V_d$ around 60. In Figure 5, the crown glass compositions BAlSi, BaNbSi, SiKZnZr, etc., explored using ML and represented as colored blobs, suggest it is possible to develop novel glasses by varying the chemical components of already explored crown glasses represented as points in the background of the blobs [42–44].To design glasses for satisfying multiple property constraints beyond the optical properties, glass selections charts [21] can be used, which are two-dimensional plots between properties of interest. Thus, the "glass veil" presented here opens up new vistas for the rational design of glasses for optical applications.

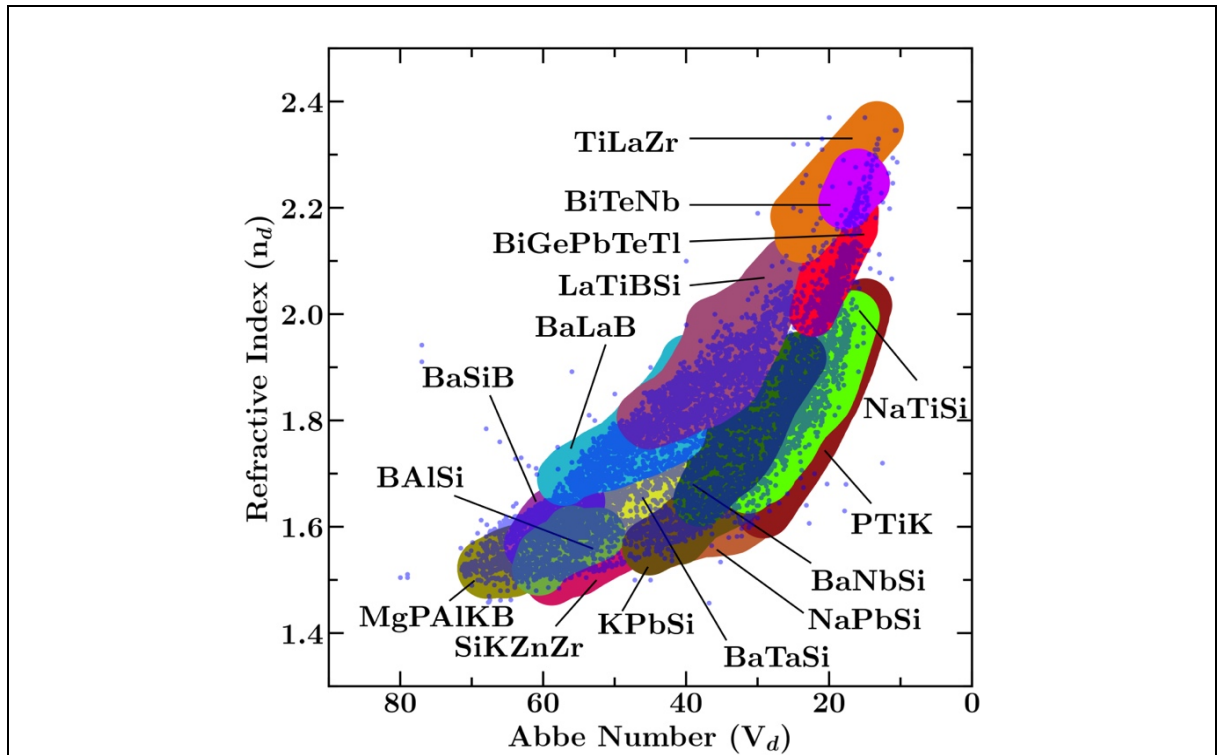

**Figure 5.** Abbe diagram (also known as the "Glass veil") generated using the ML models. Points represent experimental data. Colored regions (created using ML models) indicate the range of properties that can be achieved using respective compositions. The overlap of experimental data with ML outputs indicates the goodness of the methodology.

**Conclusion**
Overall, we demonstrated that explainable ML can be used for understanding the composition–property relationship in glasses. Specifically, we demonstrate the compositional control of two major optical properties of oxide glasses, namely, $V_d$ and $n_d$. We show that while modifiers play limited role in governing these properties, network formers and intermediates mainly control them. The approach also reveals an interesting inverse correlation where compounds that contribute in increasing the value of $V_d$ decrease the $n_d$ and vice-versa. Finally, we show that the glass veil developed using ML models allows to explore novel glass compositions beyond the pareto fronts from experiments. Altogether, the ability to interpret the black-box ML can provide deeper insights into the composition–property relationships and increase the confidence in ML-based models. More sophisticated feature selection methods may enable an exploration of the data in a more targeted manner and provide a deeper understanding of the mechanisms contributing to specific properties. This also motivates the need for newer ML methods that can better explain nonlinear dependencies of properties on composition.

**Acknowledgements**
N. M. A. K. acknowledges the financial support for this research provided by the Department of Science and Technology, India under the INSPIRE faculty scheme (DST/INSPIRE/04/2016/002774) and DST SERB Early Career Award (ECR/2018/002228). The authors thank the IIT Delhi HPC facility for providing the computational and storage resources.